\documentclass[iop]{emulateapj}
\usepackage[varg]{txfonts}
\usepackage{bm}
\usepackage{aas_macros}
\usepackage{amsfonts}
\usepackage{natbib}
\usepackage{multirow} 			
\usepackage[colorlinks,urlcolor=blue,citecolor=blue,linkcolor=blue]{hyperref}
\usepackage{enumitem}
\usepackage{cleveref} 
\usepackage{graphicx}
\usepackage{textcomp}
\usepackage[usenames, dvipsnames]{color}

\usepackage[T1]{fontenc} 
\usepackage{xhfill} 
\crefformat{footnote}{#2\footnotemark[#1]#3}

\def\src{1H\,1905+000}
\def\bfjvl{}

\def\bfpj{}
\def\bfjvlf{}
\def\bfpjf{}
\def\bfref{}
\def\bffinal{}

\def\deg{\hbox{$^\circ$}}
\def\arcmin{\hbox{$^\prime$}}
\def\arcsec{\hbox{$^{\prime\prime}$}}

\usepackage{footnote}
\usepackage{threeparttable}
\makesavenoteenv{table}

\begin{document}

\title{A search for millisecond-pulsar radio emission from the \bfpjf{faint quiescent} soft X-ray transient \src}
\shortauthors{K.~Mikhailov et al.}
\shorttitle{A search for MSP radio emission from the SXT \src}
\newcounter{aff}
\setcounter{aff}{0}

\author{
K.~Mikhailov \altaffilmark{\ref{uva},\ref{astron}},
J.~van~Leeuwen \altaffilmark{\ref{astron},\ref{uva}},
P.~G.~Jonker \altaffilmark{\ref{sron},\ref{imapp}}
}

\altaffiltext{1}{\refstepcounter{aff}\label{uva}\refstepcounter{aff}Anton Pannekoek Institute for Astronomy, University of Amsterdam, Science Park 904, PO Box 94249, 1090 GE Amsterdam, The Netherlands; \href{mailto:K.Mikhailov@uva.nl}{K.Mikhailov@uva.nl}}
\altaffiltext{\theaff}{\label{astron}\refstepcounter{aff}ASTRON, the Netherlands Institute for Radio Astronomy, PO Box 2, 7990 AA, Dwingeloo, The Netherlands}
\altaffiltext{\theaff}{\label{sron}\refstepcounter{aff}SRON, the Netherlands Institute for Space Research, Sorbonnelaan 2, 3584 CA, Utrecht, the Netherlands}
\altaffiltext{\theaff}{\label{imapp}\refstepcounter{aff}Department of Astrophysics/IMAPP, Radboud University, P.O. Box 9010, 6500 GL Nijmegen, The Netherlands}

\setcounter{aff}{0}
\setcounter{footnote}{0}

\begin{abstract}
Transitional millisecond pulsars (tMSPs) switch between an accretion-powered state \textit{without} radio pulsations and a rotation-powered state \textit{with} radio pulsations. In the former state, they are X-ray bright, in the latter X-ray dim. Soft X-ray transients (SXTs) undergo similar switches in X-ray, between ``high'' states with bright X-ray outbursts and a ``low'' state of quiescence.
The \bfpj{upper limit on the} quiescent X-ray luminosity of SXT \src\ \bfpj{suggests that its luminosity might be similar} to that of the known \mbox{tMSPs}. A detection of radio pulsations would link SXTs more strongly with \mbox{tMSPs}; and thus e.g. put stricter constraints on tMSP transitional timescales, through the connection with the well-known SXT periods of quiescence. A non-detection allows us, based on the telescope sensitivity, to estimate how likely these sources are to pulsate in radio.
Over a ten-year span, 2006-2015, we carried out targeted radio observations at 400/800\,MHz with Arecibo, and searched for radio pulsations from \bfpjf{the} quiescent SXT \src.
None of the observations have revealed radio pulsations \bfpjf{from} the targeted SXT. For a 1-ms pulsar our flux density upper limit is 10.3\,$\rm{\mu}$Jy. At an assumed distance of 10\,kpc this translates to a pseudo-luminosity \bfpjf{upper limit} of 1.0\,mJy\,kpc$^2$, which makes our search complete to $\sim85$\% of the known MSP population.
Given the high sensitivity, and the generally large beaming fraction of millisecond pulsars, we conclude that SXT \src\ is unlikely to emit in radio as a transitional millisecond pulsar.
\end{abstract}

\maketitle

\section{Introduction}\label{sec1}
\setcounter{footnote}{0}
Recent multiwavelength observations have uncovered a new type of neutron star binaries, the transitional millisecond pulsars (\mbox{tMSPs}), that switch between two separate states. The first state, an accretion-powered or low mass X-ray binary~\citep[LMXB,][]{Burderi-2013} state, is usually described by an accretion disk, formed by matter transfer from a binary companion to a neutron star (NS). It features thermal X-ray emission \bfpj{thought to arise} \bfpjf{close to} the NS surface. In this state, pulsar radio emission may be hampered \bfjvl{by e.g. surface heat (this work) or }by a disk passing through the pulsar's light cylinder~\citep{Archibald-2009}. The flat-spectrum continuum radio emission can be caused by collimated polar outflows~\citep{Deller-2015}. 

\begin{figure} 
\resizebox{\hsize}{!}
{\includegraphics{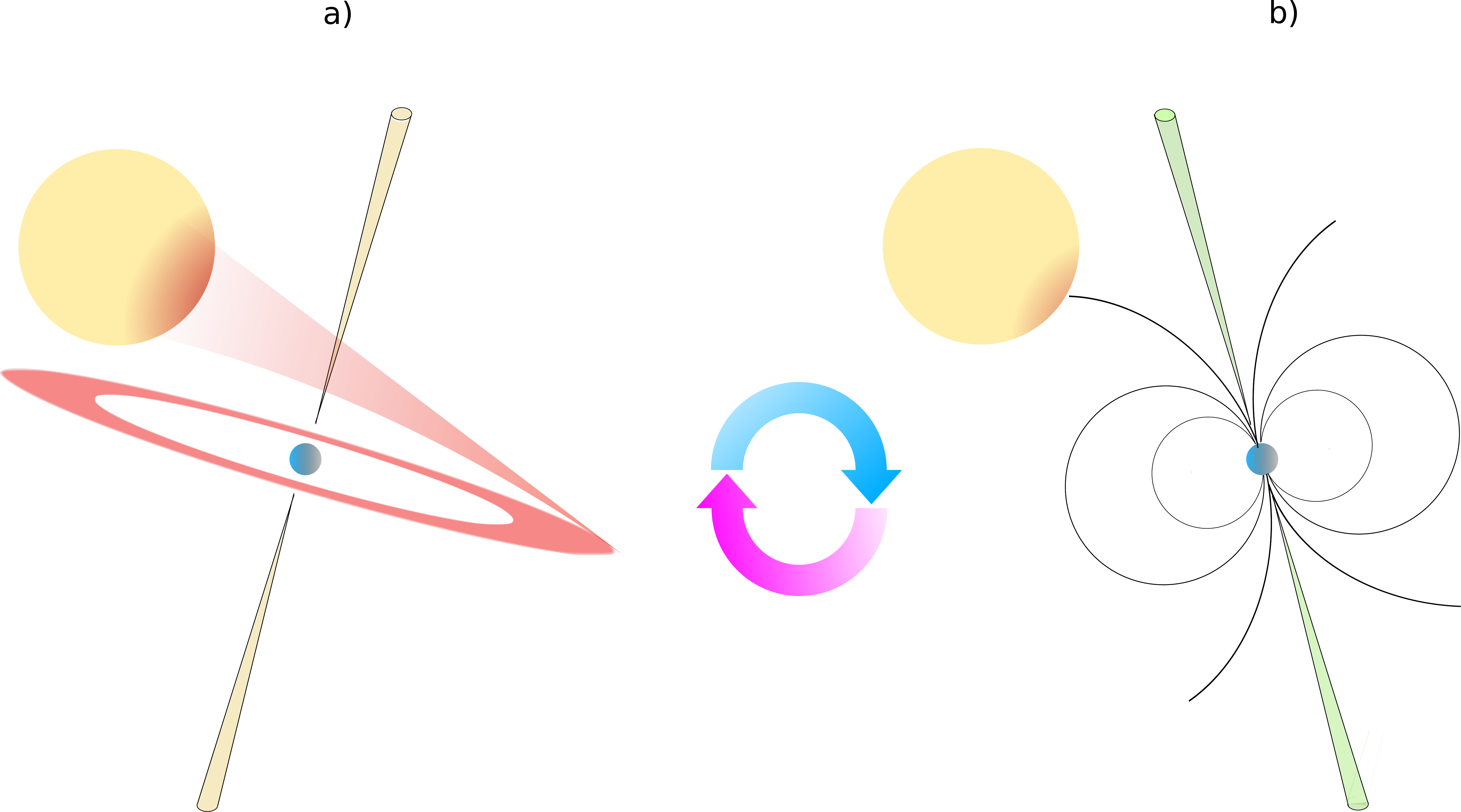}}
\caption{Two states of transitional binaries: a) an accretion-powered state with jets, collimated relativistic outflows (shown in yellow); b) a radio MSP state with a beamed radio emission (shown in green). Note that the radio pulsar beam is not necessarily in the same direction as the outflow. \bfjvlf{The red region on the companion signifies a) channelised or b) potentially on-going (but much weaker, see Sect.~\ref{ssec5.3}) accretion.}
}
\label{AMXP-MSP} 
\end{figure}

The second state, a rotation-powered or radio millisecond pulsar~\citep[MSP,][]{Bhattacharya-1991} state, lacks the surrounding disk but instead is characterised by two oppositely directed coherent radio beams from the NS polar cap region. The \bfpj{inner part of the} disk is blown away possibly by the ``propeller'' effect~\citep{Illarionov-1975} or by $\gamma$-ray photons from the pulsar magnetosphere~\citep{Takata-2014}. Both the LMXB and MSP states are illustrated in Fig.~\ref{AMXP-MSP}.

Which \bfpj{physics underlies} the tMSP switches between ``pulsar-off'' and ``pulsar-on'' states? And how do they evolve? Do all near-quiescent accreting millisecond X-ray pulsars \bfjvl{with coherent pulsations of frequency range between 150-600\,Hz}~\citep[AMXPs,][]{Patruno-2012} end up as transitional radio pulsars? \bfref{Transitions between two states might be explained via the \bffinal{interplay of} the mass accretion rate and its corresponding ram pressure, and \bffinal{the radiation pressure exerted by the pulsar wind}~\citep{Campana-1998, Burgay-2003, Archibald-2009}.}

To date, three \mbox{tMSPs} are known to undergo transitions:
\begin{enumerate}[label=(\roman*)]
\item PSR J1023+0038\bfref{~\citep{Archibald-2009}} established the tMSP class. \bfpjf{The neutron star} rotates every $1.69$\,ms and $4.8$\,hr around its rotational axis and \bfpjf{the centre of the binary mass}, respectively, and has a $\sim 0.2\,$M$_\sun$ companion;
\item PSR J1824-2452I is a binary source in the globular cluster M28~\citep[and thus henceforth referred to as M28-I,][]{Papitto-2013}, and spins every $3.93$\,ms. It also has a $\sim\,0.2\,$M$_\sun$ companion, but its orbital period is about $11$\,hr. \bfjvl{This is so far the only tMSP that also shows outbursts~\bffinal{/ Type I X-ray bursts} similar to \bfpjf{soft} X-ray transients};
\item PSR J1227-4853\bfref{~\citep{deMartino-2010, Bassa-2014, Roy-2014}}, also spins with a $1.69$\,ms period. It takes $6.7$\,hr for the source to orbit \bfpj{the center of mass}. The companion has a mass between $0.17\,$M$_\sun$ and $0.46\,$M$_\sun$.
\end{enumerate}

\bfpjf{MSPs with companions of similar mass are known as} ``redbacks''~\citep{Roberts-2013}. The MSP-state X-ray luminosity \bffinal{of all three \mbox{tMSPs}} is also of the same order: $L_\mathrm{X}\sim10^{30}$--$10^{32}$\,erg\,s$^{-1}$~\citep{Linares-2014}. \bfjvl{One tMSP (M28-I) was found \bffinal{in an X-ray bright outburst state with} $L_\mathrm{X}\sim10^{34}$--$10^{37}$\,erg\,s$^{-1}$, but \bfpjf{similar to the two other \mbox{tMSPs}} (PSR J1023+0038 and PSR J1227-4853) \bfpjf{it also exhibits} a disk-state \bffinal{at lower luminosity} $L_\mathrm{X}\sim10^{32}$--$10^{34}$\,erg\,s$^{-1}$}. The latter two also provide $\gamma$-ray emission~\citep{Stappers-2014, Johnson-2015} which establishes a putative link with a few variable $\gamma$-ray LMXBs~\citep{Strader-2015, Bogdanov-2015}. Based on multiwavelength tracking of these sources, the estimated timescale \bffinal{between disk-MSP} transitions is of order a decade~\citep{Tam-2014, Johnson-2015}. \bffinal{At the same time, transitional timescales for \mbox{tMSPs} were constrained to be much shorter -- from a few days to a few months~\citep{Stappers-2013, Papitto-2013, Bassa-2014}.}

The hypothesis we aim to test in this paper is whether \mbox{tMSPs} are related to the NS soft X-ray transients~\citep[SXTs,][]{Stella-1994}. SXTs undergo transitions between two X-ray states: a \bfpj{bright ``outburst''} state ($L_\mathrm{X}\sim10^{35}$--$10^{38}$\,erg\,s$^{-1}$)\bffinal{, sometimes} with \bfpjf{regular} bright thermonuclear flashes of type I X-ray bursts from the NS surface~\citep{Galloway-2008b}, and a much \bfpj{fainter} ``quiescent'' state ($L_\mathrm{X}\sim10^{30}$--$10^{3\rm{3}}$\,erg\,s$^{-1}$) where the source remains dim~\citep[see, e.g.][for a review]{Campana-1998, Yakovlev-2004}. \bffinal{Besides, SXTs outburst rise/decay times are also within days -- months range~\citep{Campana-1998}. Finally, the $B>10^{8}$\,G magnetic fields constrained for some X-ray transients~\citep[e.g.,][]{DiSalvo-2003} are of the same order as for AMSPs/\mbox{tMSPs}~\citep[see, e.g., Table 4 from][]{Patruno-2012}. The similarities with the X-ray luminosity, magnetic field, and transitional timescales of the tMSP makes us wonder if in this quiescent state the radio pulsar might appear.}

It is not yet clear which conditions a binary system should possess in order to undergo state switches\bfjvl{; SAX J1808.4-3658\bfref{~\citep{Wijnands-1998, Chakrabarty-1998, Hartman-2008}} is a good representation of an AMXP with about a 2.5 ms period and a typical 10$^8$-10$^9$\,G magnetic field that also \bffinal{experiences outbursts of 10$^{36}$\,erg\,s$^{-1}$ X-ray luminosity~\citep[see, e.g.,][]{Hartman-2009} but so far has not been seen in radio\bfref{~\citep{Burgay-2003, Iacolina-2010, Patruno-2016}}.} \bfpj{Given that lack of understanding on transitional-pulsar system parameters, we seek to expand the population by investigating tMSP candidates, and potentially link \bfpjf{SXTs and \mbox{tMSPs}}.}

One of the best such candidate systems is SXT \src~\citep{Jonker-2006, Jonker-2007}. It is a very low luminosity NS binary, involving a white or a brown dwarf companion ($M_\mathrm{comp}\simeq0.01-0.05\,M_\sun$) and an extremely short orbit (the orbital period $P_\mathrm{orb}\la1.33$\,hr). After about \bfpjf{an} $11$-yr ($L_X\simeq4\times10^{36}$\,erg\,s$^{-1}$) outburst \bffinal{with several Type I X-ray bursts} \bfpjf{ending} more than eighteen years ago, the source has remained quiescent. 

The quiescent luminosity of SXT \src\ $L_X\la1.8\times10^{30}$\,erg\,s$^{-1}$ is the lowest among all known NS X-ray transients, which makes the source most closely related to the MSP class. \bfref{As for hotter neutron stars, the radio pulsar emission may be inhibited via electron braking by inverse Compton scattering (see Sect.~\ref{ssec5.2} \bfref{and the references therein}), the low effective temperature $T_\mathrm{eff}\la 3.5\times10^5$\,K of SXT \src\ should allow particle to flow more freely~\citep[see, e.g.,][]{Supper-2000}. Therefore, the presumed MSP should be able to emit in radio.} Past 25\,ks~\citep{Jonker-2006} and 300\,ks~\citep{Jonker-2007} Chandra X-ray observations did not detect the NS in quiescence. No new X-ray \bfref{outbursts} were detected in X-ray all-sky monitors, and the source stayed quiescent over the period covered in this paper. This \bfref{also} suggests that SXT \src\ \bfpjf{might} be capable of producing pulsar radio emission.

\bfjvl{Revealing} radio pulsations in a quiescent SXT would link \bfjvl{the timescales on which \mbox{tMSPs} switch, to known periods of SXT quiescence. That would be a first step to better understand the interaction between accretion, thermal and radio pulsar emission.} In the remainder of this paper, Section~\ref{sec2} describes our observations of SXT \src. The data analysis for each targeted search is detailed in Sect.~\ref{sec3}. Section~\ref{sec4} outlines the obtained results, followed by the discussion in Sect.~\ref{sec5}, and conclusions in Sect.~\ref{sec6}.

\section{Observations}\label{sec2}
\bfjvl{We have conducted three targeted observations of the SXT \src\ spanning a period of almost ten years}. All three observations were performed with the 305-m William E. Gordon radio telescope at Arecibo\footnote{\url{http://www.naic.edu/index_scientific.php}}, using the L-wide (first two observations) and S-low (third observation) receivers\footnote{\url{http://www.naic.edu/~astro/RXstatus/}}. The setup specifications, detailed in Table~\ref{Obs}, are summarized below.

The first targeted search consisted of two separate observing sessions on 2006 May 20 and June 25. We used the Wideband Arecibo Pulsar Processor~\citep[WAPP\footnote{\label{wapp}\url{http://www.naic.edu/~wapp/}},][]{Dowd-2000} so that each 128\,$\mu$s it separately recorded four sets of 512 spectral frequency channels, each 0.195\,MHz wide, of 16-bit total intensity. To avoid a frequency range often affected by radio-frequency interference (RFI) we split the recording over 100 and 300\,MHz bands around 1120 and 1410\,MHz, respectively.

For our second observation on 2015 April 14 we used the Puertorican Ultimate Pulsar Processing Instrument (PUPPI) backend\footnote{\url{http://www.naic.edu/~astro/guide/node11.html}}, with 800\,MHz bandwidth and 2048 frequency channels. This resulted in the 1380\,MHz central frequency, 0.390\,MHz bandwidth per channel and the minimum sampling time of 41\,$\mu$s. 
All frequencies outside the L-wide receiver range of 1.15-1.73\,MHz were zapped during the RFI masking stage in our search pipeline.

Finally, we carried out a third observation on 2015 August 26, with the same specifications as the second but at a higher central frequency $\nu_\mathrm{obs}$ = 2800\,MHz. That reduced potential scattering on the pulsar, and was less affected by RFI.

We first pointed Arecibo on a number of known pulsars to test both telescope and backend efficiencies. Test pulsars are given in Table~\ref{test} along with respective timespans, pulse periods, dispersion measures (DMs), duty cycles, nominal flux densities, and peak signal-to-noise ratios (S/Ns). These were generally detected as predicted. The absence of PSR J1906+0746 at 2.8\,GHz was not unexpected as geodetic precession causes its radio beam to ever more miss Earth~\citep{Leeuwen-2015}.

\begin{table}
\caption{Observational setup for three different Arecibo observations of SXT \src.}
\label{Obs}
\centering
\scalebox{0.80}{
\begin{tabular}{c c c c}
\tableline\tableline
Parameter & \multicolumn{3}{c}{Value} \\
\tableline\tableline
Project ID & p2204 & p2994 & p3013 \\
Receiver (year) & L-wide (2006) & L-wide (2015) & S-low (2015) \\
Gain (K/Jy) & 10 & 10 & 8 \\
System Temp (K) & 30 & 30 & 35 \\
Native Polarization & \multicolumn{3}{c}{d u a l~~~~~~~~~~~l i n e a r} \\
Sample time ($\rm{\mu}$s) & 128 & 40.96 & 40.96 \\
Integration time (s) & 3600 & 3600 & 2150 \\
Bandwidth (MHz) & 300 & \bfjvlf{500} & \bfjvlf{500} \\
Central freq (MHz) & 1410 & 1380 & 2800 \\
Channels & 1536 & 2048 & 2048 \\
Subbands & 64 & 64 & 64 \\
\tableline\tableline
\end{tabular}
}
\tablecomments{For the first 2006 observation we list only the dataset with the larger contiguous band.}
\end{table}

\begin{table}
\caption{Parameters for test observations of two known pulsars.}
\label{test}
\centering
\begin{threeparttable}
\renewcommand{\arraystretch}{2.0}
\scalebox{0.7}{
\begin{tabular}{c c c c c c c}
\tableline\tableline
Pulsar & $t_\mathrm{int}$ & $P$ & DM & $W_{50}/P$ & $S_\nu$ & $S/N_\mathrm{peak}$ \\
	& (s) & (ms) & (pc\,cm$^{-3}$) & & (mJy) & \\
\tableline
\multirow{3}{*}{PSR J1857+0943} & \multirow{3}{*}{300} & \multirow{3}{*}{5.362} & \multirow{3}{*}{13.3} & \multirow{3}{*}{9.7$\times10^{-2}$} & \multirow{2}{*}{5.0} & 39.20~\tnote{\emph{a}} \\
								 & & & & & & 134.53~\tnote{\emph{b}} \\
								 & & & & & 1.4 & 327.35~\tnote{\emph{c}} \\
[0.5cm]
\multirow{3}{*}{PSR J1906+0746} & 600 (900)~\tnote{\emph{a}} & \multirow{3}{*}{144.073} & \multirow{3}{*}{217.20} & \multirow{3}{*}{4.2$\times10^{-3}$} & \multirow{2}{*}{0.55} & 590.64~\tnote{\emph{a}} \\
								 & 275~\tnote{\emph{b}} & & & & & 18.10~\tnote{\emph{b}} \\
								 & 900~\tnote{\emph{c}} & & & & 0.16 & (\dots)~\tnote{\emph{c}} \\
\tableline
\end{tabular}
}
\begin{tablenotes}
\footnotesize
	\item[\emph{a}]{2006 L-wide observation}
	\item[\emph{b}]{2015 L-wide observation}
	\item[\emph{c}]{2015 S-low observation}
\end{tablenotes}
\tablecomments{In 2006, PSR J1857+0943 was observed in session 1, and PSR J1906+0746 in both sessions, but with different integration times. $W_{50}/P$ is the pulse duty cycle \bfref{($W_{50}$ is the pulse width at half-peak height)}, $S_\nu$ is the pulsar flux density \bfref{based on ATNF pulsar catalogue}, and $S/N_\mathrm{peak}$ is the integrated signal-to-noise ratio at the correct DM.}
\end{threeparttable}
\end{table}

\bfref{We pointed the telescope at the best known position at the time, RA=$19{\rm^h08^m27^s}$, DEC=$+00\degr10\arcmin08\arcsec$ (J2000). This puts the Arecibo beam well over the improved position published later in~\citet{Jonker-2006}, as discussed in Sect.~\ref{ssec5.1}. We observed for} 3400\,s and 3600\,s during two first search sessions, for 3600\,s during the second search session, and for about 2150\,s during the third search session.

\section{Analysis}\label{sec3}
All our observations were analysed on the Dutch national supercomputer, Cartesius\footnote{\url{https://surfsara.nl/systems/cartesius}}. For the search itself we \bfref{adopted} the \texttt{PRESTO} pulsar search toolkit\footnote{\url{https://github.com/scottransom/presto}}~\citep{Ransom-2001}.

\subsection{2006 L-wide observation}
The WAPP backend produces multiple sequential time series, divided into 100\,MHz bandwidth slices\cref{wapp}. We used the dataset with the contiguous 300\,MHz bandwidth for the direct search, and the 100-MHz dataset only for verification. The estimated distance to the source is 7-10\,kpc~\citep{Jonker-2004}, and according to the galactic electron density model~\citep[ne2001\footnote{\url{http://www.nrl.navy.mil/rsd/RORF/ne2001/}},][]{Cordes-2002} the free electron content amounts 250-350\,pc\,cm$^{-3}$, respectively. As the ne2001 model continues to get supplemented and improved, especially for sources located far along the Galactic plane where most of cold interstellar plasma resides~\citep[see, e.g.,][]{Yusifov-2004, Cordes-2004, Sun-2010}, we searched out to a 3 $\times$ larger maximum DM, over a range from 0 to 1000\,pc\,cm$^{-3}$. \bfref{For the 80\,min orbital period and \bfref{an ultra-compact binary ($M_\mathrm{comp}\simeq0.05\,M_{\odot}$)} one should expect a \bfref{high} orbital acceleration~\citep{Johnston-1991}. Using a 0.05\,ms sampling time, we employed the maximum possible number of acceleration drift \bfref{(Fourier)} bins allowed in \texttt{PRESTO} ($\texttt{zmax} = 0, 5, 100, 200, 1200$).}

\bfref{Assuming a circular, edge-on orbit ($e = 0$, $\theta = 90\deg$), the estimated orbital acceleration is $a = (2\pi)^{4/3} \times G^{1/3} \times M_\mathrm{comp} \times \sin\theta / (P_\mathrm{orb}^{4/3} \times (M_\mathrm{ns}+M_\mathrm{comp})^{2/3})\simeq$ 28.6\,m\,s$^{-2}$, at the upper boundary of the limiting range of linear orbital accelerations $a^*\approx30$\,m\,s$^{-2}$ of most known pulsar binaries~\citep{Camilo-2000}. As the pulsar acceleration completely reverses every half of the orbit, it is reasonable to have an integration time \bffinal{within a small fraction of the orbital period~\citep[generally, $0.1\,P_\mathrm{orb}$,][]{Ransom-2003, Burgay-2003}}. For this reason, we also split up the time series to $t_\mathrm{int}\simeq410$\,s ($\simeq7$\,min) and performed the search on these shorter sets, each with less Doppler smearing. Knowing our observing duration, and aiming to be sensitive down to a period $P = 1$\,ms, we would need $N_\mathrm{drift} = a \times t_\mathrm{int}^2/(c \times P) \simeq 20$ Fourier bins. This is well in line with our covered bin range, even for a more massive companion or a faster MSP.}

\bfref{The searches over the 7-min sets have an additional sensitivity benefit if this pulsar, like other low-mass binaries, is eclipsed over part of the binary orbit~\citep[][also see Sect.~\ref{ssec5.4}]{Luo-1995}. Shorter integrations then suffer less from the addition of noise, without periodic signal, during the eclipse.}

After a number of standard periodic search steps (RFI masking, dedispersion, FFT transform, red-noise removal, RFI birdies zapping, and accelerated search; also see~\citealt{Mikhailov-2016}), all potential candidates were ranked and inspected according to their characteristic plots (pulse and frequency channels profiles, DM curve, and $P-\dot{P}$ map, see, e.g., Fig.~\ref{DM896.10_Z200}).

\begin{figure} 
\resizebox{\hsize}{!}
{\includegraphics{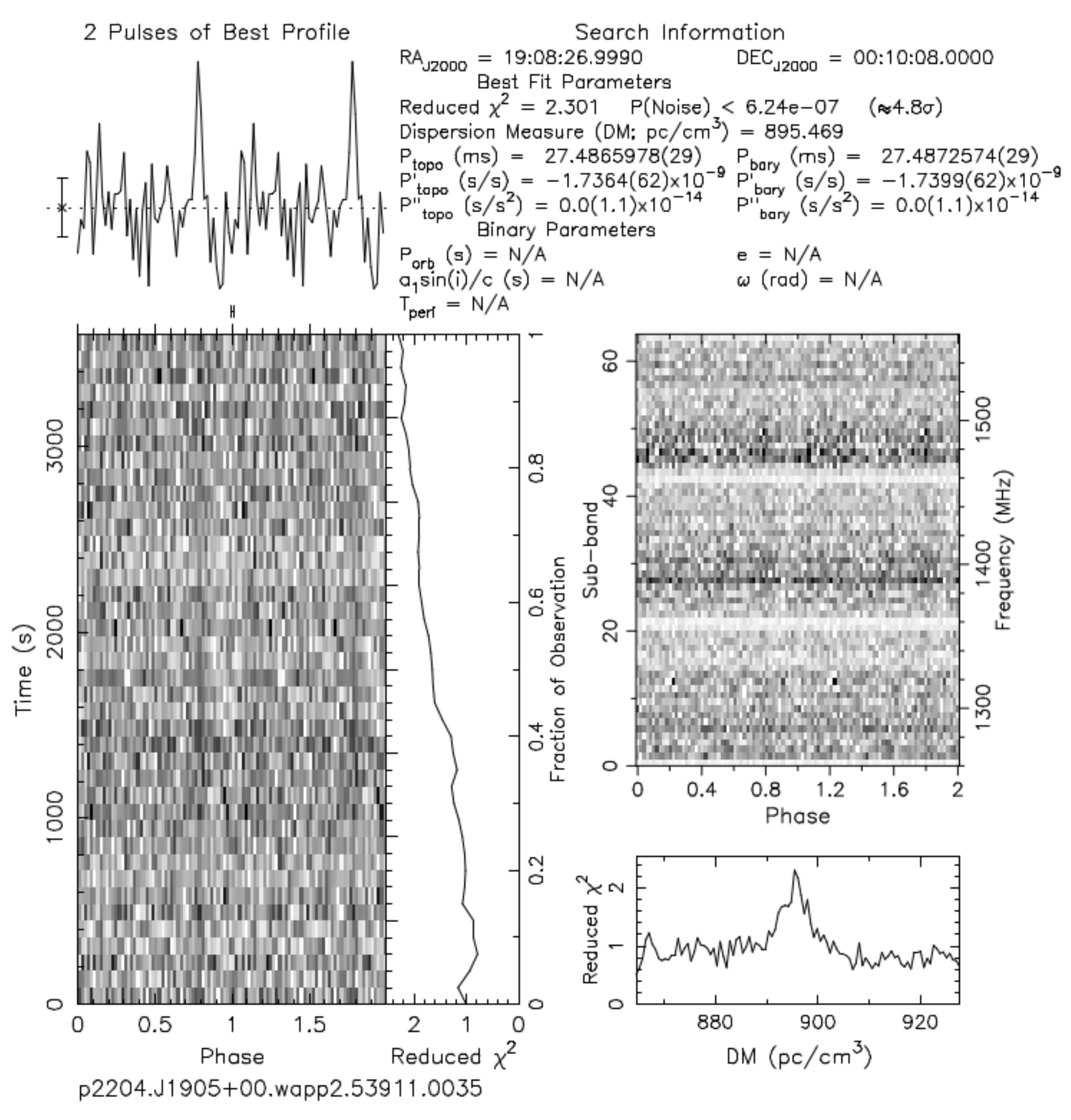}}
\caption{Representation of one of the most promising pulse configurations during the 2006 observation. Unfortunately, the candidate was only present in the second session. Moreover, further search identified that there were more similar instances for multiple DM values -- a property inherent to RFI sources.} 
\label{DM896.10_Z200} 
\end{figure}

\subsection{2015 L-wide observation}
In the 2015 1.4-GHz observations using PUPPI, the full bandwidth was recorded, including bands prone to RFI. Those were excised in the pipeline, strongly reducing the effective band width as expected. We again carried out an acceleration search but now up to 100 Fourier bins \bfref{(5 times an estimated value, see above). We did employ time-series slicing, and other search pipeline steps were also kept the same as in the 2006 observation.}

\subsection{2015 S-low observation}
In order to avoid the RFI-contaminated part of the spectrum present in the 1.4 GHz data, and to reduce the potential interstellar scattering smearing present there, we also performed a 2.8-GHz (S-low band) observation, with otherwise identical settings as the 1.4 GHz setup.

\subsection{Single-pulse search}
Some pulsars do not emit periodically, but only sporadically -- such sources are referred to as rotating radio transients~\citep[RRATs,][]{McLaughlin-2006}. For such pulsars, data folding mostly adds noise, and therefore single-pulse investigations are more effective. As millisecond pulsars emit around 10 -- 1000 pulses per second, it is hard to resolve individual pulses since their S/N is normally quite low. 

At the same time, some pulsars occasionally emit giant pulses -- a type of non-frequent short duration radio pulses that can greatly exceed (by 2 -- 3 orders of magnitude) the mean flux density of normal pulses from those pulsars. Around ten millisecond pulsars are now known to emit giant pulses~\citep{Knight-2006, Bilous-2015}, the most pronounced examples being PSR B0531+21 (the Crab), PSR B0833-45 (Vela), and PSR B1937+21. 

The most common S/N values found for candidates in our search were from 5 to 7, below \bfpj{our single-pulse} detection threshold S/N$_\mathrm{min} \geq 8$. Single-pulse thresholds can in practice be set somewhat lower than periodicity thresholds (cf.~Eq.~\ref{rad_eq}) without producing overwhelming numbers of candidates. None of the candidates above this threshold were identified as genuine pulses. Most showed a monotonic profile of S/N distribution along adjacent DM values, behaviour typical for RFI.

\section{Results}\label{sec4}
Neither the 2006 nor the 2015 observations reveal any good candidates for the potential radio MSP counterpart \bfjvl{towards \src}. 

The RFI situation at 1.4\,GHz likely impeded the identification of genuine radio pulses from \src; especially the 2015 observation, where the bandwidth increase over 2006 mostly added bands of unprotected, multi-use spectrum, contained significant amounts of RFI that resulted in the elimination of some frequency coverage, thus diminishing the potential candidate S/N compared to an RFI-free ideal.

Using the radiometer equation~\citep{Bhattacharya-1998}, we can set an upper limit on the source flux density:
\begin{equation}
S_\mathrm{min} = \beta\frac{(S/N_\mathrm{min})\,T_\mathrm{sys}}{G\sqrt{n_\mathrm{p}\,\Delta\nu\,t_\mathrm{int}}}\sqrt{\frac{W}{P-W}}.
\label{rad_eq}
\end{equation}

Here, $\beta$ is a digitisation factor and is normally around 1, $S/N_\mathrm{min}$ is the minimum distinguishable S/N from a potential pulsar, $T_\mathrm{sys}$ = $T_\mathrm{sys,receiver}$ + $T_\mathrm{sky}$ is the overall system temperature. The L-wide and S-low receiver temperatures are $T_\mathrm{sys,LW}\sim25$\,K and $T_\mathrm{sys,SL}\sim35$\,K, respectively. The sky temperature $T_\mathrm{sky}$ was interpreted from 408\,MHz contour maps~\citep{Haslam-1982} and scaled to 1400 and 2800\,MHz using the~\citet{Lawson-1987} relation $T_\mathrm{sky}\propto\nu^{-2.6}$, resulting in 5 and 0.8\,K, respectively. Next, $G_\mathrm{L-wide} = 10$\,K/Jy ($G_\mathrm{S-low} = 8$\,K/Jy) is the telescope gain\footnote{\url{http://www.naic.edu/~astro/RXstatus/}}, $n_	\mathrm{p}=2$ for two polarisations, $\Delta\nu$ is the frequency bandwidth, and $t_\mathrm{int}$ is the integration time. $P$ and $W$ are pulsar spin period and its pulse width, respectively. For typical MSP properties\footnote{\url{http://www.cv.nrao.edu/~sransom/Exascale_Radio_Searches_2014.pdf}} (spin periods $P\la20$\,ms, and magnetic field strengths $B\la10^{9}$\,G), we find\footnote{\url{http://www.atnf.csiro.au/research/pulsar/psrcat/}~\citep[catalogue version 1.54,][]{Manchester-2005}} the average pulse duty cycle $\langle W_\mathrm{50}/P \rangle$ to be 0.10\,$\pm$\,0.07. We also include smearing effects into the pulse width broadening and bound it with a pulse period: $W = W(P) = \sqrt{\langle W_\mathrm{50}/P \rangle^2 \cdot P^2 + W_\mathrm{smear}^2 + W_\mathrm{scatter}^2}$, where $W_\mathrm{smear}$ is the instrumental (channel, subband, DM$_\mathrm{step}$, \bfjvl{sampling}) smearing, and $W_\mathrm{scatter}$ is the scattering broadening along the line of sight. For the $S/N_\mathrm{min}$ we follow the usual\footnote{\url{http://www.astro.cornell.edu/~cordes/PALFA/palfa_snr_calcs.pdf}} value of 10.

\begin{table}
\caption{Estimated instrumental ($W_\mathrm{smear}$) and scattering ($W_\mathrm{scatter}$) broadening, minimum flux density $S_\mathrm{min}$, pseudo luminosity $L_\mathrm{pseudo}$ as well as search completeness $C_\mathrm{search}$ for \src\ at a distance of $d$ = 7\,kpc (DM = 250\,pc\,cm$^{-3}$) and $d$ = 10\,kpc (DM = 350\,pc\,cm$^{-3}$).}
\label{LSmin}
\centering
\setlength{\tabcolsep}{2.5pt}
\renewcommand{\arraystretch}{1.5}
\begin{tabular}{c c c c c c}
\tableline\tableline
$d$ & $W_\mathrm{smear}$ & $W_\mathrm{scatter}$ & $S_\mathrm{min}$ & $L_\mathrm{pseudo}$ & $C_\mathrm{search}$ \\
(kpcs) & (ms) & (ms) & ($\mu$Jy) & (mJy kpc$^2$) & (\%) \\ 
\tableline
\multicolumn{6}{c}{2006 L-wide observation} \\
\tableline
7 & $0.23$ & $8.4\times 10^{-3}$ & $11.8$ & $0.6$ & $89.72$ \\
10 & $0.27$ & 0.01 & $13.0$ & $1.3$ & $76.64$ \\
\tableline
\multicolumn{6}{c}{2015 L-wide observation} \\
\tableline
7 & $0.20$ & $8.5\times 10^{-3}$ & $8.5$ & $0.4$ & $91.59$ \\
10 & $0.28$ & 0.01 & $10.3$ & $1.0$ & $84.11$ \\
\tableline
\multicolumn{6}{c}{2015 S-low observation} \\
\tableline
7 & $0.06$ & $0.4\times 10^{-3}$ & $10.8$ & $0.5$ & $55.14$ \\
10 & $0.07$ & $0.6\times 10^{-3}$ & $11.1$ & $1.1$ & $31.78$ \\
\tableline
\end{tabular}
\tablecomments{All estimates are tabulated for the lowest \bfjvlf{expected} integer \bfjvlf{spin} period $P_\mathrm{min} = 1\,\mbox{ms}$.}
\end{table}

In Table~\ref{LSmin} we list our set of flux-density upper limits.
For the first 1.4\,GHz we list only the result with the longer integration time $t_\mathrm{int}=3600\,s$. The limits from the 2015 1.4\,GHz and 2.8\,GHz observations are based on 500\,MHz of usable bandwidth out of the 800\,MHz total. We see that scattering is not expected to play a significant role in the sensitivity \bfjvl{compared to other smearing effects} and that it does not change much with increasing distance. We also derive the pseudo luminosity $L_\mathrm{pseudo} = S_\mathrm{min}\,d^2$; and the search completeness $C_\mathrm{search}$, the fraction of currently known pulsars that our search could have detected at the distance of \src~\citep{Coenen-2011}.

The sensitivity of our searches is plotted in Fig.~\ref{P-S}, for various possible pulse periods and dispersion measures. In what we think is the most realistic case, a DM = 250--350\,pc\,cm$^{-3}$ (highlighted in the plot) at a pulse period of 10\,ms, our deepest search, the 2015 L-wide observation, had a minimum detectable flux of about 5.5\,$\mu$Jy.

\begin{figure} 
\resizebox{\hsize}{!}
{\includegraphics{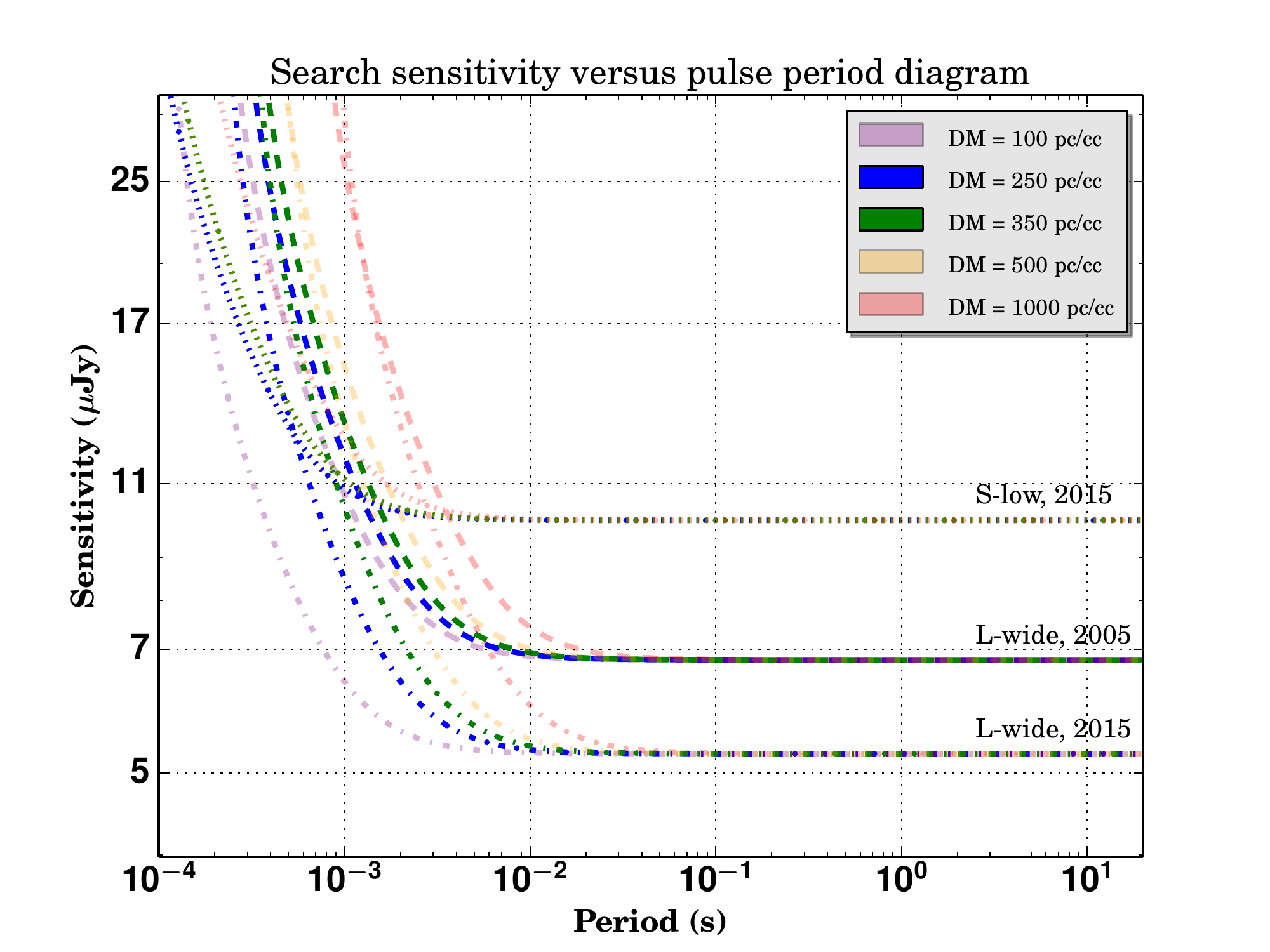}}
\caption{Sensitivity curves for each of the three \src\ Arecibo observations for source DM = 100, 250, 350, 500, and 1000\,pc\,cm$^{-3}$. Smearing effects (both instrumental and astrophysical) are included. Highlighted are the expected dispersion measure lines for $d$ = 7\,kpc (DM = 250\,pc\,cm$^{-3}$, shown in blue) and for $d$ = 10\,kpc (DM = 350\,pc\,cm$^{-3}$, shown in green).} 
\label{P-S} 
\end{figure}

The resulting search efficiency, the fraction of known pulsars whose 1.4\,GHz pseudo luminosity our survey could have detected at the distance of \src\, is shown in Fig.~\ref{Combar}. There we mark each Arecibo observation and estimated distance. To compare our 2.8\,GHz S-low observation with the 1.4\,GHz catalogue flux densities, we scaled down the search minimum detectable flux using a spectral index $\alpha = -1.8$~\citep{Maron-2000}. For each observation, we considered a pulse period range to be from 0.5\,ms to 1\,s, and the resulting spread for each observation is included in the Figure. 

Our deepest search, the 2015 L-wide observation, was sensitive enough to find \bfjvlf{up to 92\%} of all catalogued pulsars (including the three known \mbox{tMSPs}) when put at 7\,kpc. 

Given the possibility that \mbox{tMSPs} shine only intermittently, the depth at the other observation epoch is equally important. In 2005 we too would have found around 75--90\% of catalogued pulsars, and two out of three known \mbox{tMSPs}.

\bfref{To validate our estimated completeness, we also searched a longer, 2015, 1.4\,GHz dataset on PSR J1906+0746 that one of us took for long-term beam evolution studies (Desvignes et al., {\em in prep.}). At flux density $S = 2\pm1\,\mu$Jy J1906+0746 is likely to be very close to our minimum detectable flux. We used the same setup as the other 2015 observations described in the current paper. We expected, from Eq.~(\ref{rad_eq}), a S/N of about 20, given the $\sim1$\,hr integration time, duty cycle $w_\mathrm{p} = W_\mathrm{50}/P$, scattering contribution $W_\mathrm{scatter} < 0.01\,\mbox{ms}$, and dispersion smearing contributions $W_\mathrm{smear}\simeq0.2\,\mbox{ms}\simeq0.05w_\mathrm{p}$. The final pulse signal to noise obtained, for the best-fit acceleration, was 23.7, well in line with the expectation. This confirms our completeness limits are realistic.}

\begin{figure} 
\resizebox{\hsize}{!}
{\includegraphics{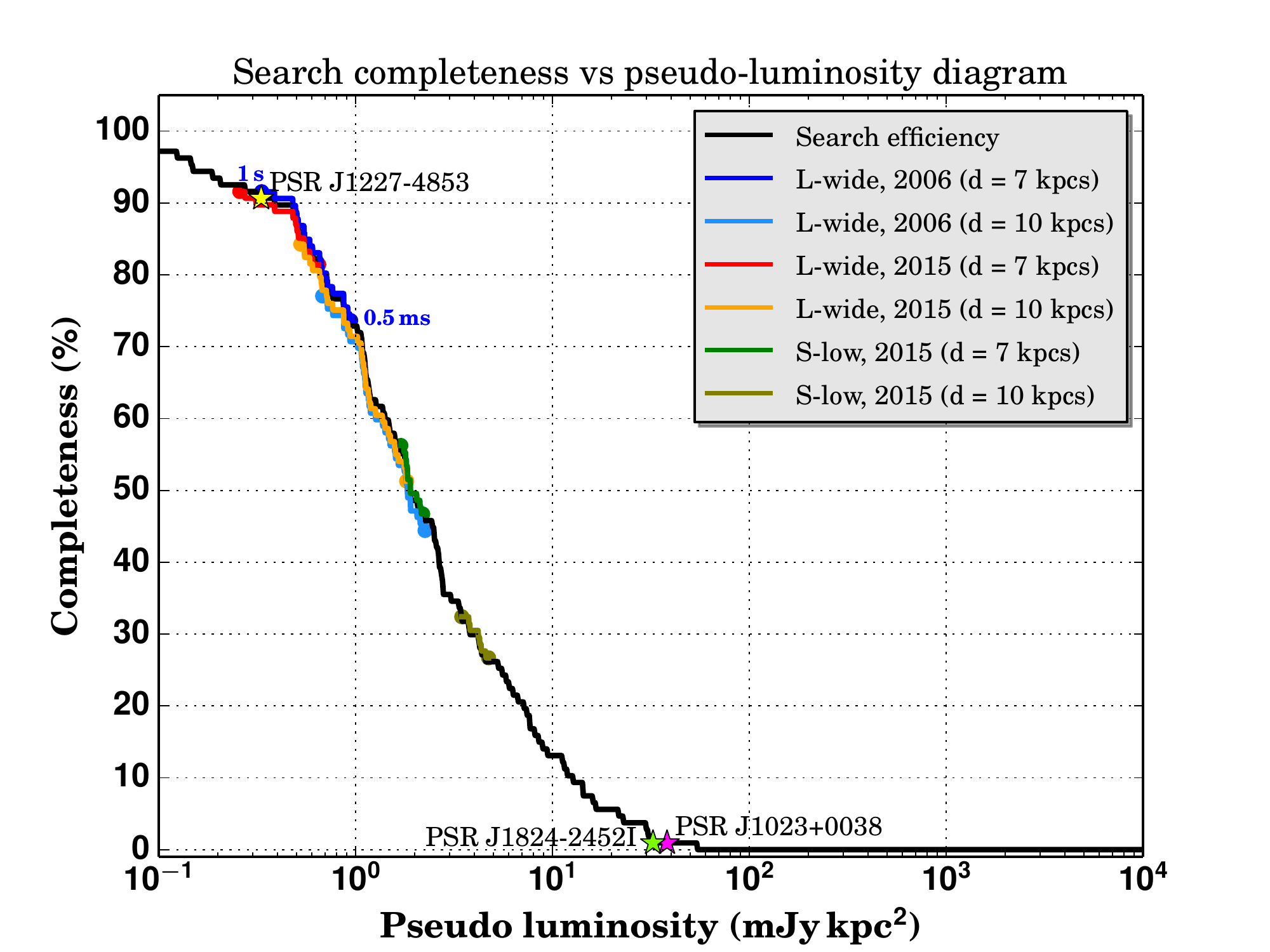}}
\caption{Search completeness compared to the ATNF-catalogue pseudo luminosities. Shown (slightly offset vertically for visibility) are the limits set by our observations. The range of values spanned by each observation is determined by the estimated pulse period, from 0.5\,ms to 1\,s (labeled as such for one example observation). The 2.8\,GHz flux densities were scaled to 1.4\,GHz for this plot.
 \bfpj{Stars represent pseudo luminosities of known \mbox{tMSPs}: J1023+0038~\citep[magenta,][]{Archibald-2009}, J1824-2452I~\citep[green,][]{Papitto-2013}, J1227-4853~\citep[yellow,][]{Roy-2014}.}} 
\label{Combar} 
\end{figure}

\section{Discussion}\label{sec5}
Given our high search completeness of $\sim$90\% we \bfpj{are confident the source is currently not a radio MSP that shines our way}. The putative pulsar itself \bfjvlf{can, of course,} in principle be inclined away from Earth. However, we know that the \bfpj{opening angle of the MSP beams} is usually quite broad, with beaming fractions of almost 100\% in several beaming models~\citep[see, e.g.,][]{Lorimer-2008}. We thus conclude that the source itself is not \bfpj{emitting radio pulses with luminosity above our limit.}

So why is \src\ not \bfpj{shining?} Below we discuss how this could be related to e.g.~surface heat, and past or current mass transfer.

\subsection{Position, proper motion and the small Arecibo beam}\label{ssec5.1}
The location of \src\ is well known. Its outburst position error box, $\alpha_{J2000.0}=19{\rm^h08^m27^s.200}\pm$0\farcs084, $\delta_{J2000.0}=+00\deg10\arcmin09\arcsec10\pm$0\farcs087~\citep{Jonker-2006}, \bfjvlf{falls well within the Arecibo FWHM beam size at both L-wide ($3.5'\times3.1'$) and S-low ($2.0'\times1.8'$) receivers.}

Could \src\ have moved out of the Arecibo beam since that outburst position was derived? Even for the smallest, S-low 2.8\,GHz beam, \src\ should then travel with proper motion $\mu = 5.4\times10^3$\,mas\,yr$^{-1}$. For the largest distance of $d=10$\,kpc, the characteristic system velocity $v_\mathrm{NS}$ needs to be $4.74\times\mu\times d\simeq 2.6\times10^5$\,km\,s$^{-1}$ for the source to have moved outside the Arecibo beam which is an order of magnitude larger than seen so far for NS X-ray binaries.

\subsection{Temperature and cooling emission}\label{ssec5.2}
In the old normal MSPs we observe, the neutron star has cooled long ago and the thermal emission is nearly absent. In transitional MSPs, some thermal emission is only seen in the accreting phase. Yet SXTs are known to be strong thermal emitters even in quiescence, due to the reheating of the neutron star by nuclear reactions deep in the crust~\citep{Brown-1998}. Models for such heating~\citep[e.g.][]{Haensel-2003, Yakovlev-2006} can explain quiescent luminosities $L_X\sim10^{33}$\,erg\,s$^{-1}$, as observed. While the limit on \src\ is much lower, as mentioned, its long and active state of accretion could still have left \bfjvlf{it} hotter than the \mbox{tMSPs}, which hardly appear \bffinal{to efficiently accrete~\citep[``radiatively inefficient'' accretion,][]{Archibald-2015, Papitto-2015, Deller-2015}} or spin up~\citep{Jaodand-2016}.

We here hypothesize that through this enhanced thermal emission over the \mbox{tMSPs}, \src\ is prevented from generating radio emission. The increased intensity of soft X-ray photons could potentially be the extinguishing agent. While some such soft thermal photons coming from the hot NS surface are needed in certain models~\citep[e.g.][]{Zhang-1997} as the seed photons that inverse Compton scatter off the primary electrons, to form pair-producing photons; these same photons, once present in larger numbers, act as radiative brakes on these essential relativistic electrons~\citep{Kardashev-1984}. The drag on these relativistic particles in the soft photon field in \src\ and its hotter brethren may thus possibly quench the runaway cascade needed for robust radio emission. 

\bffinal{If this is indeed the reason for the absence of radio emission from \src, such braking must already be significant at the temperature inferred by~\citet{Jonker-2007} of $T_\mathrm{eff}\la 3.5\times10^5$\,K, or $kT \la 30$\,eV. \citet{Supper-2000} modeled this dependence of inverse Compton scattering on temperature, for values of 100, 300, and 1000\,eV. At 100\,eV, the electrons suffer no weakening but achieve acceleration up to the maximum Lorentz factor $\gamma \simeq 10^6$. For mildly higher temperatures of 300 and 1000\,eV, respectively, electrons are quickly braked and reach Lorentz factors of 10$^4$ and 10$^2$ only (Eq.~26 and Figs. 4--6 in~\citealt{Supper-2000}), likely hampering subsequent formation of radio emission. Thus, while this model does show there is a clear dependence of  inverse Compton scattering on temperature, it cannot currently explain the case of {\src}.}

\bffinal{To summarize: it seems likely that one of the main observational differences between SXTs and \mbox{tMSPs}, the higher temperatures of the former, is the cause of the lack of radio emission on SXTs. If that is the case, braking such as inverse Compton scattering must become important at temperatures lower than proposed in \citealt{Supper-2000}, of $kT \la 30$\,eV.}

\subsection{Ongoing low-level accretion}\label{ssec5.3}
The interaction between a trickle of continuously accreting matter and the pulsar magnetosphere could potentially inhibit radio emission. There is precedent for variable low-level accretion in e.g.~Cen X-4~\citep{Cackett-2010} and XTE J1701-462~\citep{Fridriksson-2011}. A mechanism linking that accretion to the cessation of coherent radio emission was qualitatively described in \citet{Archibald-2015} to explain the X-ray bright mode in J1023+0038. One may wonder whether some low-level accretion in the X-ray dim state of \src\ remains, unnoticeable under the strict limits of $\dot{M}\lesssim10^{-13}\,M_{\sun}\,\mathrm{yr}^{-1}$~\citep{Jonker-2006}. Yet even within that limit the rate would have to be sufficient to overcome the magneto-rotational plasma, the pulsar wind, to enter the pulsar light cylinder; bringing the system into the ``propeller'' accretion regime~\citep{Illarionov-1975}. Overall we find little evidence for this explanation for the dim state of \src\ .

\subsection{Eclipses}\label{ssec5.4}
The companion mass and orbital period of \src\ are similar to those found in the ``black-widow'' (BW) MSP class~\citep{Roberts-2013}. Those pulsars and the \mbox{tMSPs} are frequently eclipsed in radio \bfref{as radio pulses are scattered by the dense wind blown off the companion due to its evaporation~\citep{Luo-1995}}. This is a possible explanation for our non-detection. \bfref{Even though all three known \mbox{tMSPs} are redbacks, similar transitions might occur in BWs. However, short exposures and relatively little photon counts from Chandra X-ray observations of nearby BW MSPs~\citep{Gentile-2014} make it difficult to prove a BW-tMSP scenario for \src\ given its X-ray upper limits.} 

\subsection{Lingering magnetic field burial}
The formation of radio-pulsar emission requires a certain magnetic field strength. Yet accretion is thought to bury and diminish radio-pulsar magnetic fields during the transition from dead, normal pulsar to reborn \mbox{tMSPs}. Potentially, the magnetic field strength in \src\ is now, after a prolonged period of accretion, too low to power the acceleration of primary charged particles.

Accretion rates above $\dot{M} \simeq 0.03\,\dot{m}_\mathrm{Edd}$, where $\dot{m}_\mathrm{Edd}$ is the Eddington accretion rate, are required for diamagnetic screening (``burial'') to be efficient~\citep{Cumming-2001}. The active-state mass accretion rate estimated to be required for 11 years of outbursts, $\dot{M}\simeq 10^{-9}\,M_{\sun}\,\mathrm{yr}^{-1} \simeq
0.1\,\dot{m}_\mathrm{Edd}$~\citep{Jonker-2006} is larger and could thus bring about such screening. \bffinal{In that accretion steady state, the magnetic field was likely diminished by $n \simeq (\dot{M} / 0.02\,\dot{m}_\mathrm{Edd}) = 5$ orders of magnitude~\citep{Cumming-2008}. In the most extreme outburst modelled in~\citet{Cumming-2008}, of 2\,yr at $0.05\,\dot{m}_\mathrm{Edd}$, the magnetic field re-emerges on a $\sim$1\,yr timescale (Fig.~2 in \citealt{Cumming-2008} for a $^{28}$Si ocean). The significantly longer and more intense (11\,yr, $0.1\,\dot{m}_\mathrm{Edd}$) outburst of \src\ may possibly be causing lingering magnetic field burial on the observed $\sim$10\,yr timescale, preventing radio pulsations.}

Given the quiescent mass transfer rate of $10^{-13}\,M_{\sun}\,\mathrm{yr}^{-1}$ it might take around $10^4$ yr to build up a new accretion disk of the same amount of mass~\citep{Jonker-2006}, and start a new active phase. Thus even in high-accretion rate systems such as \src\, the magnetic field appears to have time to re-emerge, and effectuate radio-pulsar emission.

\subsection{A low-magnetic field neutron star}\label{ssec5.6}
\bfref{Finally, as the accretion onto a NS lowers its magnetic field, large amounts of accreted matter may at some point demolish the electrical currents that supply NS magnetisation~\citep{Shibazaki-1989, Romani-1990, Miri-1994}. This can ultimately lead to a very large Ohmic magnetic field decay, resulting in a NS with an extremely low magnetic field~\citep[see, e.g.,][]{Tauris-2001}.} In regular pulsars such a long-term effect\bfref{~\citep{Sang-1987}}, reducing the magnetic field by four orders of magnitude, is well-known (and illustrated in Fig.~\ref{ppdot}). But as MSPs, like normal pulsars, can only shine above some ``death line''~\citep{Chen-1993, Zhang-2000}, there is strong selection effect toward MSPs with magnetic field of at least 10$^8$\,G. If \src\ due to its accretion history now supports only an extremely low-strength \bffinal{down to $B\simeq10^7$\,G~\citep{Zhang-2013}} field, it would remain radio quiet. In this case, the NS would likely have had to accrete many tenths of solar masses overall, without collapsing to a black hole, as a significant number of MSPs that have gained similar weight in the past are still shining today~\citep{Antoniadis-2016}. \bffinal{If this is the reason \src\ is not shining, it should be positioned in a region of the standard $P-\dot{P}$ diagram (see Fig.~\ref{ppdot}) that is above its limiting magnetic field line, but below death line~\citep[e.g.,][]{Zhang-2000}, closest to the actual MSP population but as yet invisible.}

\begin{figure}[t!]
\resizebox{\hsize}{!}
{\includegraphics{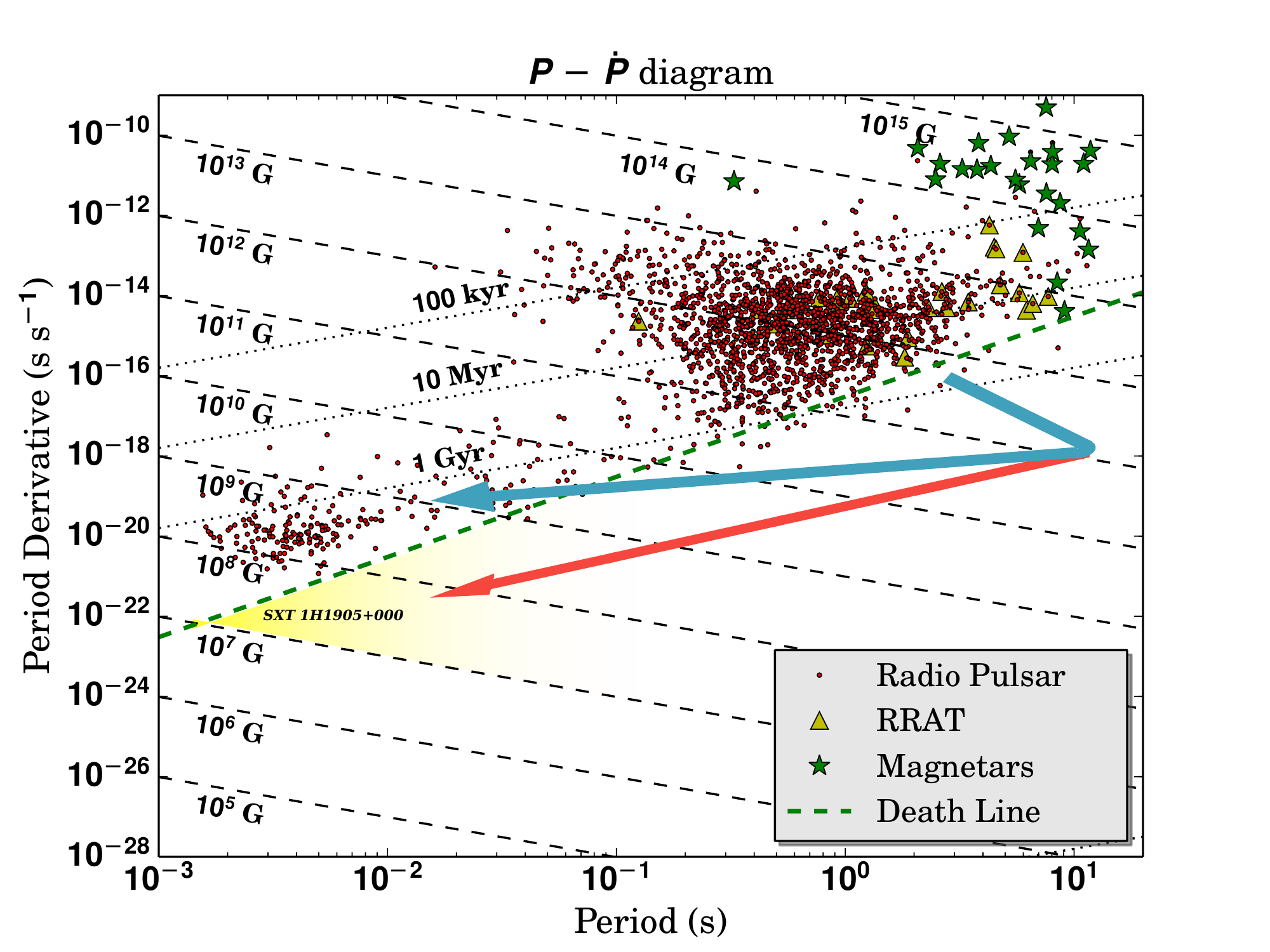}}
\caption{Two evolutionary scenarios for radio millisecond pulsars on a general $P-\dot{P}$ diagram: a standard ``recycling'' evolution~\citep[][blue line]{Bhattacharya-1991} and a low magnetic field ``reconfiguration'' evolution (red line). \bffinal{Only the area segment between the limiting $B\simeq10^7$\,G line and the closest to known MSPs death line~\citep[green dashed line, see Fig. 1 from][line III$\arcmin$]{Zhang-2000} is suitable for SXT \src. The brightness of the yellow area represents the positional probability of the transient.}}
\label{ppdot} 
\end{figure}

\section{Conclusion and Future Work}\label{sec6}
Our Arecibo observations toward the dimmest known soft X-ray transient, \src, did not reveal pulsar emission in over a decade. We have set a strong limit on its pseudo luminosity at largest distance $L_\mathrm{pseudo\,(d = 10\,kpc)}$ = 1.0\,mJy\,kpc$^2$. \bfpj{We are 85\% confident that SXT \src\ is presently not in the radio pulsar state.} Future, more sensitive and simultaneous radio and X-ray observations might reveal the exact nature of SXTs and put more constraints on their potentially tight connection with the traditional MSP class. Additional gamma-ray observations could potentially shed light on the radio-beaming fraction and the radio-eclipse scenarios, and thus support more firmly our current conclusion: \src\ is inherently radio quiet. 

\section*{Acknowledgements}
We thank L.~Bildsten, S.~Ransom, G.~Nelemans and I.~Stairs for the discussions that started this project, and A. Archibald, R. Wijnands\bffinal{, and the referee} for subsequent suggestions. The Arecibo Observatory is operated by SRI International under a cooperative agreement with the National Science Foundation (AST-1100968), and in alliance with Ana G. M\'endez-Universidad Metropolitana, and the Universities Space Research Association. The research leading to these results has received funding from the European Research Council under the European Union's Seventh Framework Programme (FP/2007-2013) / ERC Grant Agreement n. 617199, and from the Netherlands Research School for Astronomy (NOVA4-ARTS). We also thank SURF Cooperative and e-Science support center for provision and maintenance of the Dutch national supercomputer Cartesius and related resources. Computing time was provided by NWO Physical Sciences.

\bibliography{/home/klim/Dropbox/PhD/LaTeX/Ref}{}
\bibliographystyle{yahapj}
\end{document}